\documentclass[showpacs,aps,pra,twocolumn]{revtex4}
\usepackage{amsmath,graphicx,bbm,mathrsfs,amssymb,pst-all,bm,color}

\input{tcilatex}
\begin{document}

\title{Quantum algorithm for obtaining the energy spectrum of a physical
system}
\author{Hefeng Wang$^{1, 2, 3}$, S. Ashhab$^{2, 3}$, and Franco Nori$^{2, 3}$%
}
\affiliation{$^{1}$Department of Applied Physics, Xi'an Jiaotong University, Xi'an
710049, China\\
$^{2}$Advanced Science Institute, RIKEN, Wako-shi, Saitama 351-0198, Japan\\
$^{3}$Department of Physics, The University of Michigan, Ann Arbor, Michigan
48109-1040, USA}

\begin{abstract}
We present a polynomial-time quantum algorithm for obtaining the energy
spectrum of a physical system, i.e. the differences between the eigenvalues
of the system's Hamiltonian, provided that the spectrum of interest contains
at most a polynomially increasing number of energy levels. A probe qubit is
coupled to a quantum register that represents the system of interest such
that the probe exhibits a dynamical response only when it is resonant with a
transition in the system. By varying the probe's frequency and the
system-probe coupling operator, any desired part of the energy spectrum can
be obtained. The algorithm can also be used to deterministically prepare any
energy eigenstate. As an example, we have simulated running the algorithm
and obtained the energy spectrum of the water molecule.

\noindent
\end{abstract}

\maketitle

\section{Introduction}

Obtaining the energy spectrum of a physical system is an important task in a
variety of fields. In general, one has to solve the Schr\"{o}dinger equation
of the system, which is a difficult task on a classical computer for large
systems, because the dimension of the Hilbert space of the system increases
exponentially with the size of the system, which is commonly defined as the
number of particles in the system. Thus, the complexity of simulating the
quantum system grows exponentially. On a quantum computer, however, the
number of qubits required to simulate the system increases linearly with the
size of the system. As a result, Solving the Schr\"{o}dinger equation of the
system is more efficient on a quantum computer than on a classical computer~%
\cite{bacon,mhy,jybb, nori1, nori2}.

The standard quantum algorithm for obtaining the eigenvalues and
eigenvectors of the Hamiltonian of a quantum system is the phase estimation
algorithm~(PEA)~\cite{kitaev95, abrams, aa, whf, whf0, childs, nori3, whf2}.
In the PEA, one prepares an initial guess state, and the algorithm randomly
\textquotedblleft selects\textquotedblright\ one of the energy eigenstates
in the guess state and produces its energy as the output of the algorithm.
It is worth mentioning here that the probability of selecting a given energy
eigenstate is equal to the square of its overlap with the guess state. In
reality, one is usually most interested in the energy differences between
energy levels, instead of the absolute energy of a given energy level. In
this paper, we present a quantum algorithm that solves this problem:
obtaining the energy differences between energy levels of a quantum system
described by a given Hamiltonian. The algorithm can also be used to prepare
any energy eigenstate of the system.

Our algorithm is motivated by the following observation in simulating the
dynamics of an open quantum system~\cite{terhal, whf1, sanders}: For an open
system interacting with many environment modes, the mode that resonates with
a certain transition in the spectrum of the open system contributes the most
to the decay dynamics associated with that transition. This property
suggests a method to locate the transition frequencies separating the
different energy levels of a physical system.

The basic idea of the algorithm is as follows: we couple the quantum system
to a probe qubit with a certain frequency, set the probe qubit in one of its
energy eigenstates~(say the excited state), evolve the whole system for some
time, then perform a measurement on the probe qubit. When the frequency of
the probe qubit matches the transition frequency between two energy levels
of the quantum system, one observes a peak in the decay rate of the probe
qubit. Therefore by varying the frequency of the probe qubit, we can locate
the transition frequencies of the quantum system. We can also set the probe
qubit to be in its ground state and measure its excitation dynamics. The
difference is that in the former case we obtain the absorption spectroscopy
of the system while in the latter case we obtain the emission spectroscopy.

This algorithm has the following advantages: $\left( i\right) $ There are
several adjustable elements~(initial state of the system, interaction
operator, evolution time and system-probe coupling strength) that can be
varied in order to improve the efficiency of the algorithm. $(ii)$ The
coupling of the system to the probe qubit can simulate a realistic
interaction, and therefore the algorithm can naturally identify transitions
that would occur in a realistic setting. $(iii)$ Because of the freedom
associated with choosing the coupling operator, the algorithm gives as an
additional piece of output the transition matrix elements for any desired
operator. $(iv)$ Because the algorithm involves transitions between
different energy eigenstates, preparing the system in a good approximation
to any particular energy eigenstate is less crucial than in the phase
estimation algorithm.

The structure of this work is as follows: In Sec.~\ref{alg}, we present an
algorithm for obtaining the energy spectrum of a physical system. In Sec.~%
\ref{example}, we give an example to demonstrate the algorithm for obtaining
the energy spectrum of the water molecule. In Sec.~\ref{discuss}, we discuss
the efficiency, the accuracy and the resource requirement of the algorithm,
and compare our algorithm with the PEA. We close with a conclusion section.

\section{The algorithm}

\label{alg}

First, we make an initial guess about the range of the energy differences
between the energy levels of the system, $\left[ \omega _{\min }\text{, }%
\omega _{\max }\right] $. We discretize this frequency range into $j$
intervals, where each interval has a width of $\Delta \omega =\left( \omega
_{\max }-\omega _{\min }\right) /j$, and the center frequencies are given by
$\omega _{k}=\omega _{\min }+\left( k+1/2\right) \Delta \omega ,k=0\ldots
,j-1$. We now let a probe qubit couple to the quantum system, and we design
the Hamiltonian of the whole system to be of the form%
\begin{equation}
H=H_{S}+\frac{1}{2}\omega \sigma _{z}+cA\otimes \sigma _{x},
\end{equation}%
where the first term is the Hamiltonian of the system, the second term is
the Hamiltonian of the probe qubit, and the third term describes the
interaction between the system and the probe qubit. Here, $\omega $ is the
frequency of the probe qubit~(we have set $\hbar=1$), and $c$ is the
coupling strength between the probe qubit and the system, while $\sigma _{x}$
and $\sigma _{z}$ are Pauli matrices. The operator $A$ acts in the state
space of the system and plays the role of an excitation operator that
transfers the initial state of the quantum system to another state. The
frequency $\omega $ is taken from the frequency set $\omega _{k}$. For a
frequency $\omega _{k}$ of the probe qubit, we let the whole system evolve
with the Hamiltonian shown in Eq.~($1$) for a time $\tau $. This evolution
is implemented using the procedure of quantum simulation based on the
Trotter-Suzuki formula~\cite{nc}. After that, we read out the state of the
probe qubit. We repeat the whole procedure many times in order to obtain the
decay probability. Then we change the probe frequency and repeat this
procedure until we cover all the frequencies in the range $\left[ \omega
_{\min }\text{, }\omega _{\max }\right] $.

Setting the probe qubit in its excited~(ground) state, when the frequency of
the probe qubit matches the transition frequency between two energy levels
of the quantum system, the probe qubit has the fastest decay~(excitation).
For example, in the case where the initial state of the probe qubit is the
excited state, the final state of the probe qubit is:
\begin{equation}
\rho _{p}(\tau )=\text{Tr}_{S}[U(\tau )\left( |\psi _{s}\rangle \langle \psi
_{s}|\otimes |1\rangle \langle 1|\right) U^{\dag }(\tau )],
\end{equation}%
where Tr$_{S}[\cdots ]$ means tracing out the system degrees of freedom. The
unitary evolution operator $U(\tau )=\exp \left( -iH\tau \right) $, $H$ is
given in Eq.~($1$), $|\psi _{s}\rangle $ is the initial state of the system,
and $|1\rangle $ represents the excited state of the probe qubit, while $%
|0\rangle $ represents the ground state of the probe qubit. The quantity of
interest to us now is the decay probability of the probe qubit $P_{\text{%
decay}}=\langle 0|\rho _{p}(\tau )|0\rangle $. By plotting $P_{\text{decay}}$
as a function of the probe-qubit frequency, we can obtain the absorption
spectrum of the system. If there are no degeneracies in the transition
frequencies, at most one transition~(denoted by $i\rightarrow j$) in the
system will contribute to the decay dynamics of the probe qubit~(taking into
consideration the possibility of degenerate transitions makes the
derivations longer but does not affect our main results). In this case, we
obtain the result
\begin{equation}
P_{\text{decay}}=\sin ^{2}\left( \frac{\Omega _{ij}\tau }{2}\right) \frac{%
Q_{ij}^{2}}{Q_{ij}^{2}+\left( E_{j}-E_{i}-\omega _{k}\right) ^{2}}|\langle
\varphi _{i}|\psi _{s}\rangle |^{2},
\end{equation}%
where $Q_{ij}=2c|\langle \varphi _{j}|A|\varphi _{i}\rangle |$, and $\Omega
_{ij}=\sqrt{Q_{ij}^{2}+\left( E_{j}-E_{i}-\omega _{k}\right) ^{2}}$. $%
|\varphi _{i}\rangle $ is the $i$-th energy eigenstate of the system and $%
E_{i}$ is the corresponding eigenenergy. Eq.~($3$) describes
Rabi-oscillation dynamics, where the system and probe exchange an
excitation. The second factor on the right-hand side is the maximum
oscillation probability, and it depends on the relation between the matrix
element for a given transition and the system-probe detuning for that
transition. The third factor is the overlap between the initial state of the
system and a given energy eigenstate.

In general, the interaction between the probe qubit and the quantum system
should be weak such that the widths of the peaks are small and one obtains
accurate results. The evolution time $\tau $ should ideally be large~($c\tau
\sim 1$), such that the change of the system is clear and the peaks in the
spectroscopy have high resolution.

The procedure of the algorithm is as follows: $\left( i\right) $ prepare a
quantum register $R_{S}$, which encodes the state of the system, in state $%
|\psi _{s}\rangle $, and the probe qubit in state $|1\rangle $; $\left(
ii\right) $ implement the unitary operator $U(\tau )=\exp \left( -iH\tau
\right) $ where $H$ is given in Eq.~($1$); $\left( iii\right) $ read out the
state of the probe qubit; $\left( iv\right) $ perform steps $\left( i\right)
$ -- $\left( iii\right) $ many times in order to obtain good statistics and
calculate the decay probability; $\left( v\right) $ repeat steps $\left(
i\right) $ -- $\left( iv\right) $ for different frequencies of the probe
qubit. From the above procedure, one obtains the absorption spectroscopy of
the system. One can also set the probe qubit in its ground state $|0\rangle $%
, and perform the above steps in order to obtain the emission spectroscopy.
The quantum circuit for steps $\left( i\right) $ -- $\left( iii\right) $ is
shown in Fig.~$1$.

\begin{figure}[tbp]
\includegraphics[width=0.9\columnwidth, clip]{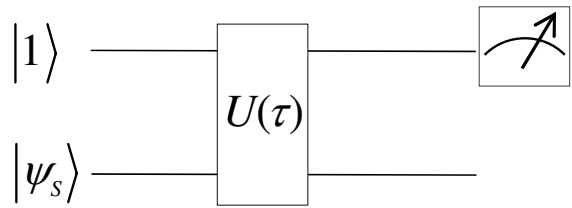}
\caption{Quantum circuit for obtaining the energy spectrum of a physical
system. The first input register represents a probe qubit, and the second
input register represents the system whose spectrum we are trying to obtain.}
\end{figure}

\section{Example: obtaining the energy spectrum of the water molecule}

\label{example}

In the following, we present an example that demonstrates how the algorithm
would perform in obtaining the energy spectrum of the water molecule.

To apply the quantum algorithm presented above, first we have to map the
state space of the water molecule to the state space of the qubits. Using
the mapping technique introduced in Ref.~\cite{whf}, considering the C$_{2V}$
and $^{1}A_{1}$ symmetries known from quantum chemistry of the water
molecule, we can minimize the number of qubits needed to represent the water
molecule on a quantum register. Note that the symmetries can be used with
the PEA in order to optimize the algorithm in the same way that we have used
them in the water-molecule example. These symmetries would not lead to an
exponential speedup in either algorithm. In other words, knowing the
symmetries is not crucial for running the algorithm and for having a
polynomial scaling of resources. The Hamiltonian for the water molecule is
given in Ref.~\cite{sbo} and shown below. The Hartree-Fock wave function for
the ground state of the water molecule is $%
(1a_{1})^{2}(2a_{1})^{2}(1b_{2})^{2}(3a_{1})^{2}(1b_{1})^{2}$. Using the STO-%
$3$G basis set~\cite{sbo} and freezing the first two $a_{1}$ orbitals, we
construct a model space with $^{1}A_{1}$ symmetry that includes the $%
3a_{1},4a_{1},1b_{1}$ and $1b_{2}$ orbitals and we consider only single and
double excitations to the external space for performing the
multi-reference-configuration interaction~(MRCI) calculation. The MRCI space
is composed of $18$ configuration state functions. Therefore at least $5$
qubits are required to represent the state of the water molecule in this
calculation.

In order to optimize the implementation of the algorithm, it is useful to
have a priori knowledge of the molecular states and their symmetries. This
can be done using quantum-chemistry algorithms on a classical computer.

The Hamiltonian of the water molecule in the form of second quantization is

\begin{equation}
H=\sum_{p,q}\left\langle p\left\vert T+V_{N}\right\vert q\right\rangle
a_{p}^{\dagger }a_{q}-\frac{1}{2}\sum\limits_{p,q,r,s}\left\langle
p\left\vert \left\langle q\left\vert V_{e}\right\vert r\right\rangle
\right\vert s\right\rangle a_{p}^{\dagger }a_{q}^{\dagger }a_{r}a_{s},
\end{equation}%
where $|p\rangle $ is the one-particle state, $a_{p}^{\dagger }$ is its
fermionic creation operator, and $T$, $V_{N}$, and $V_{e}$ are the
one-particle kinetic operator, nuclear attraction operator and the
two-particle electron repulsion operator, respectively.

For the initial state, we prepare the system register in the simple state $%
|00010\rangle $, which is close to the true ground state. Then we implement
the unitary operation $U=\exp \left( -iH\tau \right) $. For the interaction
operator $A$, we set
\begin{equation}
A=(A_{1}+A_{2}+A_{3}+A_{4}+A_{5})/\sqrt{5},
\end{equation}%
where $A_{1}=I\otimes I\otimes I\otimes I\otimes \sigma _{x}$, $%
A_{2}=I\otimes I\otimes I\otimes \sigma _{x}\otimes I$, $A_{3}=I\otimes
I\otimes \sigma _{x}\otimes I\otimes I$, $A_{4}=I\otimes \sigma _{x}\otimes
I\otimes I\otimes I$, $A_{5}=\sigma _{x}\otimes I\otimes I\otimes I\otimes I$%
. We set the coupling strength $c=0.005$ and the evolution time $\tau =500$
(here we measure energies in units of Hartree and time in units of Hartree$%
^{-1}$). We vary the frequency of the probe qubit in the range $\omega \in %
\left[ 0.4\text{, }2.0\right] $, which is divided into $200$ intervals, and
run the circuit shown in Fig.~$1$. We obtain the spectrum shown in Fig.~$2$
for the transition frequencies between the ground state and several excited
states. From the figure we can see that the spectroscopy obtained using our
algorithm is in very good agreement with the known transition frequency
spectrum~(in red) of the water molecule.

The coupling strength $c$ and the evolution time $\tau $ can be adjusted to
improve the resolution of the peaks and the accuracy of the results. In
order to demonstrate this point, we now set $c=0.001$ and $\tau =2500$. We
focus on the second and the third peaks as shown in the inset of Fig.~$2$.
We can see that the widths of the peaks are reduced and the resolution of
the peaks is now higher. We also observe a small peak at the frequency of
the transition between the second and the eighth energy levels.

From Fig.~$2$, we can see that some transitions between the ground state and
the excited states are barely visible. Their decay probabilities can be
improved by constructing a different operator $A$. The choice for the
operator $A$ in Eq.($5$) includes single-qubit operators with all the qubits
represented. With this choice most of the desired resonance peaks are
observed in the simulation. However, as can be seen in Fig.$2$, some peaks
are very low. We use two-qubit operators in order to look for any such
\textquotedblleft missing\textquotedblright\ peaks. We have tried a few
different choices, and we only show the one that resulted in all the peaks
being visible. In principle, even if no single operator (as happened in our
example) produces all the resonance peaks, one can still construct the
spectrum by putting together the information obtained from the different
choices for $A$. We define the operators $A_{6}=I\otimes I\otimes I\otimes
\sigma _{x}\otimes \sigma _{x}$ and $A_{7}=I\otimes I\otimes \sigma
_{x}\otimes \sigma _{x}\otimes I$, set the interaction operator%
\begin{equation}
A=(A_{1}+A_{2}+A_{3}+A_{6}+A_{7})/\sqrt{5},
\end{equation}%
and run the algorithm. The results are shown in Fig.~$3$. We can see that
now all the expected peaks are clearly visible.
\begin{figure}[tbp]
\includegraphics[width=0.9\columnwidth, clip]{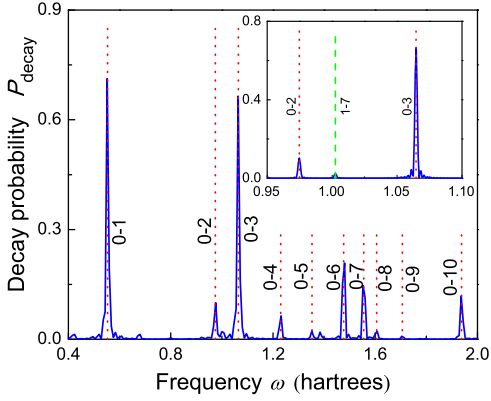}
\caption{(Color online)~Transition frequency spectrum between the ground
state, $|\protect\varphi _{0}\rangle $, and the first ten excited states $%
\left( |\protect\varphi _{i}\rangle ,i=1,2,\ldots ,10\right) $ of the water
molecule. The blue solid curve represents the decay probability of the probe
qubit at different frequencies with the coupling coefficient in Eq.~($1$) $%
c=0.005$ and the evolution time $\protect\tau =500$, and the operator $A$ as
shown in Eq.~($5$). The red dotted vertical lines represent the known
transition frequencies between the ground state and the first ten excited
states of the water molecule. In the inset, the second and the third
transition frequencies shown in blue were obtained using $c=10^{-3}$ and $%
\protect\tau =2500$. The green vertical dashed line represents the known
transition frequency ($1$-$7$) between the first and the seventh excited
states.}
\end{figure}
\begin{figure}[tbp]
\includegraphics[width=0.9\columnwidth, clip]{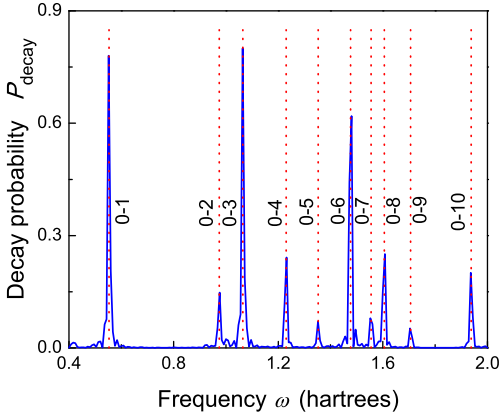}
\caption{(Color online)~Same as in Fig.~$2$, except the operator $A$ is set
as shown in Eq.~($6$).}
\end{figure}

In our algorithm, we can transfer the initial state of the system to another
state through the interaction operator $A$. Therefore the initial state of
the system is not of crucial importance to the success of the algorithm.
Here we give an example that demonstrates how the PEA can fail when the
initial state is not a good approximation of the desired state, but where
our algorithm still succeeds.

In the PEA, the success probability of the algorithm depends on the overlap
of the initial guess state with the desired eigenstate of the system. In the
previous example, if the initial state of the water molecule is prepared in
state $|11111\rangle $, the overlap of this state with any of the $18$
eigenstates~(in our example, the dimension of the state space of the water
molecule is $18$) of the water molecule is \emph{zero}. Therefore the PEA
will \emph{fail} in such a case. Our algorithm, however, still works.

We set the operator $A$ to be%
\begin{equation}
A=\frac{1}{3}\sum_{i=1}^{9}B_{i},
\end{equation}%
where $B_{1}=\sigma _{x}\otimes \sigma _{x}\otimes \sigma _{x}\otimes \sigma
_{x}\otimes I$, $B_{2}=\sigma _{x}\otimes \sigma _{x}\otimes \sigma
_{x}\otimes I\otimes \sigma _{x}$, $B_{3}=\sigma _{x}\otimes \sigma
_{x}\otimes I\otimes \sigma _{x}\otimes \sigma _{x}$, $B_{4}=\sigma
_{x}\otimes I\otimes \sigma _{x}\otimes \sigma _{x}\otimes \sigma _{x}$, $%
B_{5}=\sigma _{x}\otimes \sigma _{x}\otimes \sigma _{x}\otimes I\otimes I$, $%
B_{6}=\sigma _{x}\otimes I\otimes I\otimes \sigma _{x}\otimes \sigma _{x}$, $%
B_{7}=\sigma _{x}\otimes I\otimes \sigma _{x}\otimes I\otimes I$, $%
B_{8}=\sigma _{x}\otimes I\otimes I\otimes I\otimes \sigma _{x}$, $%
B_{9}=\sigma _{x}\otimes \sigma _{x}\otimes \sigma _{x}\otimes \sigma
_{x}\otimes \sigma _{x}$. And we use the state $|11111\rangle $ as the
initial state of the system. We set the coupling coefficient $c=0.002$, and
the evolution time $\tau =800$, and run the algorithm. The results are shown
in Fig.~$4$. From this figure we can see that the algorithm still has a high
success probability in obtaining the energy spectrum of the water molecule.
\begin{figure}[tbp]
\includegraphics[width=0.9\columnwidth, clip]{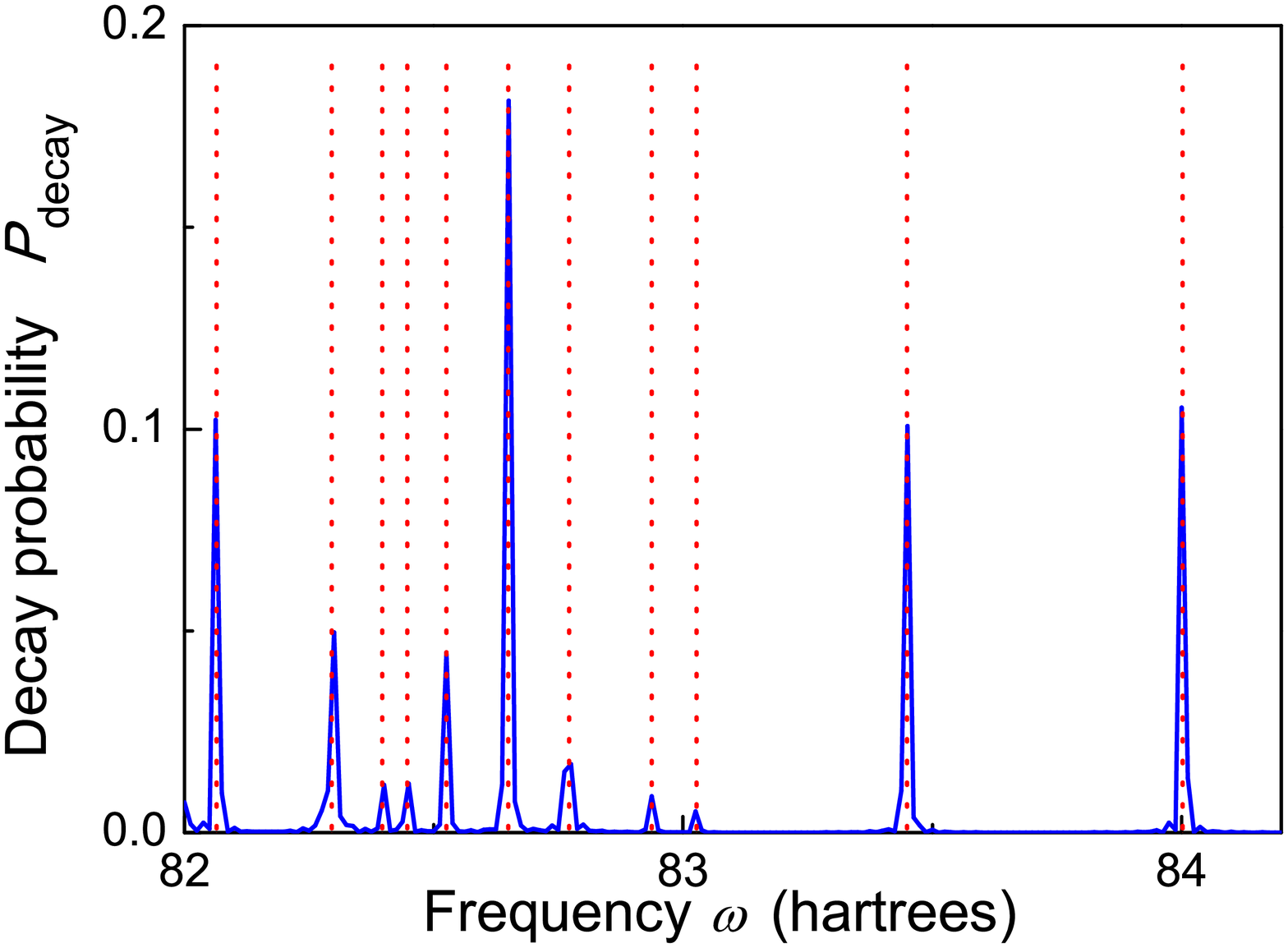}
\caption{(Color online)~Transition frequency spectrum between the first $11$
eigenstates of the water molecule and the state $|11111\rangle $. The blue
solid curve represents the decay of the probe qubit at different frequencies
with the coupling coefficient shown in Eq.~($1$) $c=0.002$ and the evolution
time $\protect\tau =800$ and operator $A$ as shown in Eq.~($7$), in
simulating the algorithm, and the red dotted vertical line represents the
known eigenenergies of the first $11$ eigenstates of the water molecule. }
\end{figure}

\section{Discussion}

\label{discuss}

In the following, we discuss the factors that affect the efficiency, the
accuracy and the resource requirements of the algorithm, and we compare our
algorithm with the phase estimation algorithm.

The efficiency of the algorithm is most naturally defined through the number
of times that the circuit in Fig.~$1$ must be run in order to identify the
peaks in the spectrum. This number is proportional to the number of
frequency points that need to be used and the number of times that the
circuit needs to be run for a single frequency. Most physical systems have
typical energy scales that are linear in the system size~(for the total
energy), while some unusual systems exhibit a polynomial dependence with
relatively small exponents. The energy scale for the low-energy spectrum
might be even smaller than that scale. The number of frequency points that
need to be used in the algorithm, which is proportional to the frequency
range, therefore scales polynomially with the system size. The number of
times that the circuit needs to be run for a single frequency must be at
least proportional to $1/P_{\text{decay}}$ in order to observe a peak. It
should also be mentioned here that each single run of the algorithm is
essentially a quantum simulation of the dynamics, which scales polynomially
with the size of the system~\cite{kitaev2}.

We note here that, for large systems, there is an exponentially large number
of energy eigenstates, and determining the entire spectrum of a large system
exhibits exponential complexity. However, one is usually not interested in
all of the energy eigenstates, but rather a very small fraction of them,
that is, a polynomial number of energy eigenstates. The energy eigenstates
of interest could for example be the low-lying energy levels or the energy
levels that are connected with the ground state by strong electric-dipole
transitions. Once the criterion for the \textquotedblleft energy levels of
interest\textquotedblright\ is specified, and their number is small~(or at
least not exponentially large), the complexity of the algorithm does not
grow exponentially with the size of the system any more. The part of the
spectrum of interest will appear naturally in our algorithm, because the
system will undergo transitions that mimic those of the simulated system.

Since the heights of the peaks~depend on the product $\Omega _{ij}\tau $, we
assume as a \textquotedblleft worst-case scenario\textquotedblright\ that $%
Q_{ij}\tau \ll 1$ and find that the decay probability in Eq.~($3$) at the
center of a given peak~(i.e. at $\omega _{k}=E_{j}-E_{i}$) can be
approximated as
\begin{equation}
P_{\text{decay}}\approx c^{2}\tau ^{2}|\langle \varphi _{j}|A|\varphi
_{i}\rangle |^{2}|\langle \varphi _{i}|\psi _{s}\rangle |^{2}.
\end{equation}%
From the above equation, we can see that the decay probability, and
therefore the efficiency of the algorithm, depends on the coupling strength,
the evolution time, the interaction operator and the initial state of the
system. Note that we can also use a number of qubits in parallel as probe
qubits in order to improve the efficiency of the algorithm. We now address
the roles of the different factors appearing in Eq.~($6$).

The parameters $c$ and $\tau $ define the accuracy of the algorithm. The
accuracy is given by the width of the peaks, and this width is given by $%
\max \left[ c\langle \varphi _{j}|A|\varphi _{i}\rangle \text{, }1/\tau %
\right] $~\cite{whf1}. To obtain accurate results, we need to set $c$ to be
small so that we have weak system-probe coupling and the evolution time $%
\tau $ is set to be large such that the change of the system remains
observable. Note that the accuracy is set by the experimenter, independently
of the size of the system. It is also worth noting here that the size of the
frequency intervals $\Delta \omega $ is set by the choice of $c$ and $\tau $%
: $\Delta \omega $ should be smaller than the width of the peaks in order to
avoid missing some of the peaks, but there is no point in reducing $\Delta
\omega $ far beyond this point.

Since $P_{\text{decay}}$ depends on $\langle \varphi _{j}|A|\varphi
_{i}\rangle $, the algorithm can also be used to obtain the matrix element $%
\langle \varphi _{j}|A|\varphi _{i}\rangle $ for any operator $A$ and any
two energy eigenstates, provided that this matrix element is not
exponentially small. For this purpose, it would be ideal to set the initial
state to one of the states $|\varphi _{i}\rangle $, which can be achieved as
will be explained shortly. One can then use the height of the peak to obtain
$\langle \varphi _{j}|A|\varphi _{i}\rangle $. In this context it is also
worth noting that since the height of the peak depends on $\langle \varphi
_{j}|A|\varphi _{i}\rangle $, certain transitions might not result in any
peaks if the relevant matrix element vanishes. By designing $A$ to be a
physically-relevant operator, e.g. the operator that describes the coupling
of a molecule to an electric field, one can identify transitions that would
occur under electromagnetic irradiation of a molecule. Needless to say, $A$
is not restricted to be a naturally occurring operator.

The last factor in Eq.~($6$) is the overlap between the initial state and
any given energy eigenstate~(which serves as the initial state in a given
transition). In principle, preparing an initial state that has a large
overlap with any given energy eigenstate can be a difficult task, possibly
involving exponential scaling in the size of the system. However, a crucial
point here is that once we observe a transition at the end of given run of
the algorithm, we know that the final state of the system is the final
energy eigenstate of the relevant transition. We can now use this state as
the new initial state and rerun the algorithm. If the new initial state is
different from a state that we wish to examine, we can convert the state of
the system into the desired state by adding or subtracting from the system
the required energy difference, which we would know at least approximately~%
\cite{nakazato}. Even if the two states in question were macroscopically
different, it should be possible to go from one of them to the other via an
at-most polynomially large number of transitions each of which involves
single- or few-body operators. In many cases of practical interest, a
relatively small number of transitions are needed to connect the energy
levels of interest, e.g. the low-lying energy eigenstates.

We note here that, since our algorithm relies on changes in the energy to
keep track of the state of the system, one cannot tell whether a given
energy level is degenerate or not, and in the case of degeneracy one cannot
tell which final state is obtained upon detecting the relevant transition.
If one wishes to check for degeneracies, one could add a few small
perturbations to the Hamiltonian of the system, and for most physical
systems these perturbations will lift the degeneracies in the spectrum.

Finally, we compare our algorithm with the PEA. In the PEA, one prepares an
initial state that ideally has a large overlap with the desired energy
eigenstate, and the algorithm produces the energy of that state. In cases
where the desired energy eigenstate has a complicated form or whose form is
unknown, it can become impossible to prepare a guess state that has any
substantial overlap with the desired state. The algorithm would fail in this
case. In our algorithm, the initial state does not need to have a large
overlap with any particular state. As mentioned above, the observation of a
transition in the probe signals a corresponding transition in the system.
The post-transition state can now be treated as the initial state for the
next step in the algorithm. This way, one can guide the system to any energy
eigenstate, including the ground state. The freedom in choosing the operator
$A$ allows additional controllability for this purpose.

We note here that there is no single choice of the operator $A$ that is
needed in order to obtain a certain energy difference, say between the
ground state and first-excited state. As explained above, it should be
possible to go from any state to any other state via a relatively small
number of intermediate states. An exception might be glassy and similar
frustrated systems with a large number of vastly different low-energy
states. However, there is no known efficient algorithm, classical or
quantum, for exhaustively identifying the low-energy states of such complex
systems.

In the PEA, one needs to have a good idea about the form of the energy
eigenstates of interest. In our algorithm no such a priori knowledge is
needed. If one is interested in the low-energy spectrum, the relevant states
would show up naturally in the spectrum. This property is demonstrated in
the example presented in Sec.~\ref{example}, showing that our algorithm can work in
cases where the PEA fails.

In the PEA, one obtains the absolute eigenenergy of the system. For a large
system, the absolute eigenenergy could be a large number, much larger than
the separation between the energy eigenstates of interest. This large
overall energy would appear as part of the output, thus taking up resources
such as additional index qubits. In our algorithm, one obtains the energy
difference between energy levels, therefore avoiding the unnecessary readout
of any overall energy shift. Note that the number of qubits required for
implementing our algorithm is the same as that in the optimized version of
the PEA~\cite{aa, griffith}.

\section{Conclusion}

We have presented a hybrid analogue/digital quantum algorithm for obtaining
the energy spectrum of a physical system. The algorithm provides more
flexibility than the phase estimation algorithm. It can also be used to
simulate a realistic interaction, and naturally identify transitions that
would occur in a realistic setting. The algorithm can also be used to
prepare any desired energy eigenstate of a physical system.

\begin{acknowledgements}
We acknowledge partial support from DARPA, AFOSR, ARO, NSF
grant No.~0726909, JSPS-RFBR contract No.~12-02-92100,
Grant-in-Aid for Scientific Research~(S), MEXT Kakenhi on Quantum
Cybernetics, and the JSPS FIRST program. HW is supported
by \textquotedblleft the Fundamental
Research Funds for the Central Universities\textquotedblright\ of China.
\end{acknowledgements}


\begin{thebibliography}{99}
\bibitem{bacon} D. Bacon and W. van Dam, Communications of the ACM, \textbf{%
53}, 84~(2010).

\bibitem{mhy} M.-H. Yung, J.~D. Whitefield, S. Boixo, D.~G. Tempel and A.
Aspuru-Guzik, e-print arXiv:1203.1331 (2012).

\bibitem{jybb} J. Yepez and B. Boghosian, Comp. Phys. Comm. \textbf{146},
280-294 (2002).

\bibitem{nori1} I. Buluta, F. Nori, Science \textbf{326}, 108 (2009).

\bibitem{nori2} I. Buluta, S. Ashhab, F. Nori, Rep. Prog. Phys. \textbf{74},
104401 (2011).

\bibitem{kitaev95} A.~Y. Kitaev, e-print arXiv: quant-ph/9707021, 1997.

\bibitem{abrams} D.~S. Abrams and S. Lloyd, Phys. Rev. Lett. \textbf{83},
5162~(1999).

\bibitem{aa} A. Aspuru-Guzik, A.~D. Dutoi, P.~J. Love, and M. Head-Gordon,
Science \textbf{309}, 1704~(2005).

\bibitem{whf} H. Wang, S. Kais, A. Aspuru-Guzik, and M.~R. Hoffmann, Phys.
Chem. Chem. Phys. \textbf{10}, 5388~(2008).

\bibitem{whf0} H. Wang, L.-A. Wu, Y.-X. Liu, F. Nori, Phys. Rev. A, \textbf{%
82}, 062303~(2010).

\bibitem{childs} A.~M. Childs and W.~van Dam, Rev. Mod. Phys. \textbf{82},
1~(2010).

\bibitem{nori3} L.F. Wei, F. Nori, J. of Phys. A \textbf{37}, 4607 (2004).

\bibitem{whf2} H. Wang, L.-A. Wu, Y.-X. Liu, F. Nori, Phys. Rev. A \textbf{82%
}, 062303 (2010).

\bibitem{terhal} B.~M. Terhal and D.~P. DiVincenzo, Phys. Rev. A, \textbf{61}%
, 022301~(2000).

\bibitem{whf1} H. Wang, S. Ashhab and F. Nori, Phys. Rev. A, \textbf{83},
062317~(2011).

\bibitem{sanders} N. Wiebe, D.~W. Berry, P. Hoyer, B.~C. Sanders, e-print
arXiv:1011.3489~(2011).

\bibitem{nc} M. Nielsen, I. Chuang, \emph{Quantum Computation and Quantum
Information}~(Cambridge Univ. Press, Cambridge 2000).

\bibitem{sbo} A. Szabo, N. Ostlund, {\normalsize \textit{Modern Quantum
Chemistry: Introduction to advanced Electronic Structure Theory}}%
~(McGraw-Hill, New York, 1989).

\bibitem{kitaev2} A. Kitaev, A. H. Shen, and M. N. Vyalyi. \emph{Classical
and quantum computation, Graduate studies in Mathematics} Vol. \textbf{47}.
American Mathematical Society, Providence, RI, 2002.

\bibitem{nakazato} H. Nakazato, T. Takazawa, and K. Yuasa, Phys. Rev. Lett.
\textbf{90}, 060401 (2003).

\bibitem{griffith} R. B. Griffiths and C.-S. Niu, Phys. Rev. Lett. \textbf{76%
}, 3228~(1996).
\end{thebibliography}
\end{document}